\documentclass[letter]{jpsj3}
\usepackage{color}
\usepackage{ulem}

\title{Interplay of Superconductivity and Fermi-Liquid Transport in Rh-Doped CaFe$_2$As$_2$ with Lattice-Collapse Transition}

\author{
\name{Masataka \surname{Danura}}$^{1,2}$, 
\name{Kazutaka \surname{Kudo}}$^{1,2}$\thanks{E-mail address: kudo@science.okayama-u.ac.jp}, 
\name{Yoshihiro \surname{Oshiro}}$^{1}$, 
\name{Shingo \surname{Araki}}$^{1}$, \\
\name{Tatsuo C. \surname{Kobayashi}}$^{1}$, 
and \name{Minoru \surname{Nohara}}$^{1,2}$
}

\inst{
$^{1}$Department of Physics, Okayama University, Okayama 700-8530, Japan \\
$^{2}$JST, Transformative Research-Project on Iron Pnictides (TRIP), Chiyoda, Tokyo 102-0075, Japan \\
}

\abst{
Ca(Fe$_{1-x}$Rh$_x$)$_2$As$_2$ undergoes successive phase transitions with increasing Rh doping in the $T$ $=$ 0 limit. 
The antiferromagnetic-metal phase with orthorhombic structure at 0.00 $\le$ $x$ $\le$ 0.020 is driven to a superconducting phase with  uncollapsed-tetragonal (ucT) structure at 0.020 $\le$ $x$ $\le$ 0.024; a non-superconducting collapsed-tetragonal (cT) phase takes over at $x$ $\geq$ 0.024. 
The breakdown of Fermi-liquid transport is observed in the ucT phase above $T_{\rm c}$. 
In the adjacent cT phase, Fermi-liquid transport is restored along with a disappearance of superconductivity. 
This interplay of superconductivity and Fermi-liquid transport suggests the essential role of magnetic fluctuations in the emergence of superconductivity in doped CaFe$_2$As$_2$. 
}

\kword{Iron-based superconductors, Lattice collapse transition, Fermi-liquid transport}

\begin{document}
\maketitle

Unconventional superconductivity often emerges in close proximity to a magnetically ordered state\cite{rf:Dagotto}. 
Uncovering the character of the magnetic fluctuations is crucial for understanding the mechanism of superconductivity. 
The magnetic fluctuations often manifest themselves in non-Fermi-liquid behavior at the magnetic quantum critical points (QCPs), points at which a magnetic phase transition takes place at the $T$ $=$ 0 limit\cite{rf:Stewart}. 
Quadratic temperature dependence of resistivity, $\rho(T)$ $\propto$ $T^2$, the hallmark of a Fermi liquid, is suppressed by the presence of low-lying magnetic fluctuations near QCPs; a non-Fermi-liquid ground state is realized at QCPs. 
Unlike in the case of cuprate or heavy fermion superconductors\cite{rf:Dagotto}, the role of magnetic fluctuations in superconductivity is still not well understood in iron-based materials\cite{rf:Ishida,rf:Paglione,rf:Johnston}. 
However, a number of experiments have indicated the presence of magnetic fluctuations in the normal state of iron-based superconductors: temperature-dependent NMR rates $(T_1T)^{-1}$ [\citen{rf:Fukazawa,rf:Mukuda,rf:Ning,rf:Nakai}] and the breakdown of Fermi-liquid transport, signaled by the non-quadratic temperature dependence of resistivity\cite{rf:Liu1,rf:Gooch,rf:Kasahara2,rf:Kasahara}. 
A $T$-linear resistivity has been widely observed in optimally doped materials such as SmFeAsO$_{1-x}$F$_x$\cite{rf:Liu1}, (Sr$_{1-x}$K$_x$)Fe$_2$As$_2$\cite{rf:Gooch}, and BaFe$_2$(As$_{1-x}$P$_x$)$_2$\cite{rf:Kasahara2}.

CaFe$_2$As$_2$ is a unique material that experiences a first-order phase transition from the uncollapsed-tetragonal (ucT) phase to the collapsed-tetragonal (cT) phase under either an applied hydrostatic pressure\cite{rf:Kreyssig,rf:Goldman} or chemical substitutions\cite{rf:Kasahara}. 
This transition is characterized by the shrinkage of the $c$ axis by approximately 10\% without breaking any symmetries. 
A band calculation predicted the lifting of the Fermi surface nesting\cite{rf:Kasahara,rf:Tompsett} along with the disappearance of the iron magnetic moment in the cT phase\cite{rf:Yildirim}.
This gives us a unique opportunity to investigate the relation between magnetism and superconductivity. 
Recently, Kasahara {\it et al.} have demonstrated that following the lattice-collapse transition in CaFe$_2$(As$_{1-x}$P$_x$)$_2$, Fermi-liquid transport abruptly recovers along with the abrupt disappearance of superconductivity\cite{rf:Kasahara}. 
Their result suggests that the magnetic fluctuations, related to Fermi-surface nesting\cite{rf:Ding,rf:Liu,rf:Brouet,rf:Terashima}, play an important role in superconductivity.

In this study, we demonstrate that the lattice collapse transition takes place in Ca(Fe$_{1-x}$Rh$_x$)$_2$As$_2$: the ucT phase at 0.020 $\le$ $x$ $\le$ 0.024; the cT phase at $x$ $\geq$ 0.024 in the $T$ $=$ 0 limit. 
Application of hydrostatic pressure also results in a lattice collapse transition at $P$ $\sim$ 0.5 GPa in Ca(Fe$_{1-x}$Rh$_x$)$_2$As$_2$ with $x$ $=$ 0.021. 
In both cases, superconductivity abruptly disappears when the ucT phase changes into the cT phase; non-Fermi-liquid transport in the ucT phase changes into Fermi-liquid transport in the cT phase. 
Moreover, we show that the $T$-linear paramagnetic susceptibility in the ucT phase changes into (approximately $T$-independent) Pauli paramagnetic behavior in the cT phase. 
All of these observations suggest the conclusion that the magnetic fluctuation in the ucT phase is lifted by the lattice collapse transitions.

Single crystals of Ca(Fe$_{1-x}$Rh$_x$)$_2$As$_2$ with 0.00 $\le$ $x$ $\le$ 0.41 were grown using a self-flux method. 
A mixture with a ratio of Ca : FeAs : RhAs = 1 : $4-4x$ : $4x$ was placed in an alumina crucible and sealed in an evacuated quartz tube. 
The ampule was heated at 1100 $^\circ$C and cooled to 1050 $^\circ$C at a rate of 1.25 $^\circ$C/h; it was thereafter allowed to cool to room temperature. 
Crystals were mechanically isolated from the flux. 
Note that a quench from 1050 $^\circ$C (to allow the centrifuge to remove flux) results in a severe degradation of the superconducting properties of the sample, this is likely caused by inhomogeneity. 
The Rhodium content $x$ was determined using an energy dispersive X-ray (EDX) analyzer. 
Lattice parameters at room temperature were determined using powder X-ray diffraction (XRD). 
Magnetization $M$ was measured using a SQUID magnetometer (Magnetic Property Measurement System, Quantum Design). 
Electrical resistivity $\rho_{ab}$ was measured using the standard DC four-terminal method with a Physical Property Measurement System (PPMS, Quantum Design). 
Measurements under hydrostatic pressure were performed using an indenter cell\cite{rf:Kobayashi}.

\begin{figure}[t]
\begin{center}
\includegraphics[width=8.5cm]{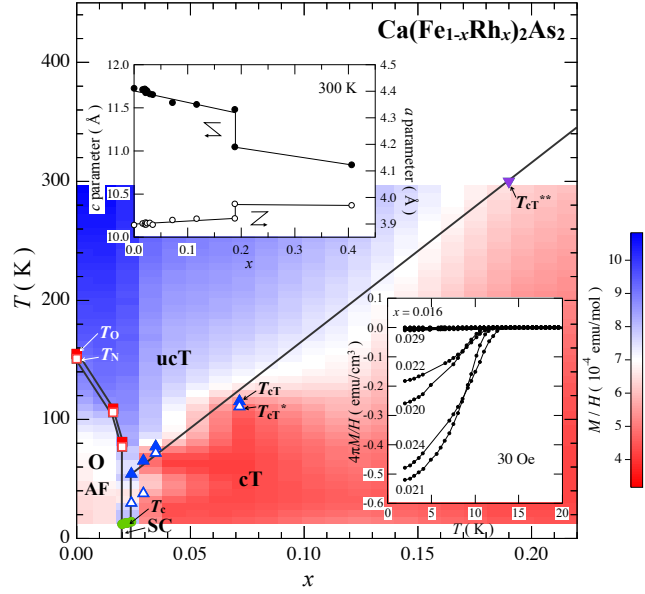}
\caption{(Color online) The electronic phase diagram of Ca(Fe$_{1-x}$Rh$_x$)$_2$As$_2$. 
O, ucT, and cT denote the orthorhombic, uncollapsed-tetragonal, and collapsed-tetragonal structure, respectively. 
SC and AF denote the superconducting and the antiferromagnetically ordered phases, respectively. 
$T_{\rm c}$ is the superconducting transition temperature, as determined from the magnetization. 
$T_{\rm O}$ and $T_{\rm N}$ are the ucT-O and AF transition temperatures, respectively, as determined from the electrical resistivity $\rho_{ab}$. 
$T_{\rm cT}$, $T_{\rm cT}^*$, and $T_{\rm cT}^{**}$ are the ucT-cT transition temperatures, as determined from $\rho_{ab}$ upon heating and cooling, and the lattice parameters at room temperature, respectively. 
The upper inset shows $x$ dependence of lattice parameters $a$ and $c$ at room temperature. 
The lower inset shows the temperature dependence of magnetization $M$ divided by the magnetic field $H$, $M/H$, in a magnetic field of 30 Oe under zero-field cooling and field cooling.
The contour plot shows the temperature and $x$ evolution of $M/H$ measured in a magnetic field of 5 T. 
} 
\end{center}
\end{figure}
\begin{figure}[t]
\begin{center}
\includegraphics[width=8.5cm]{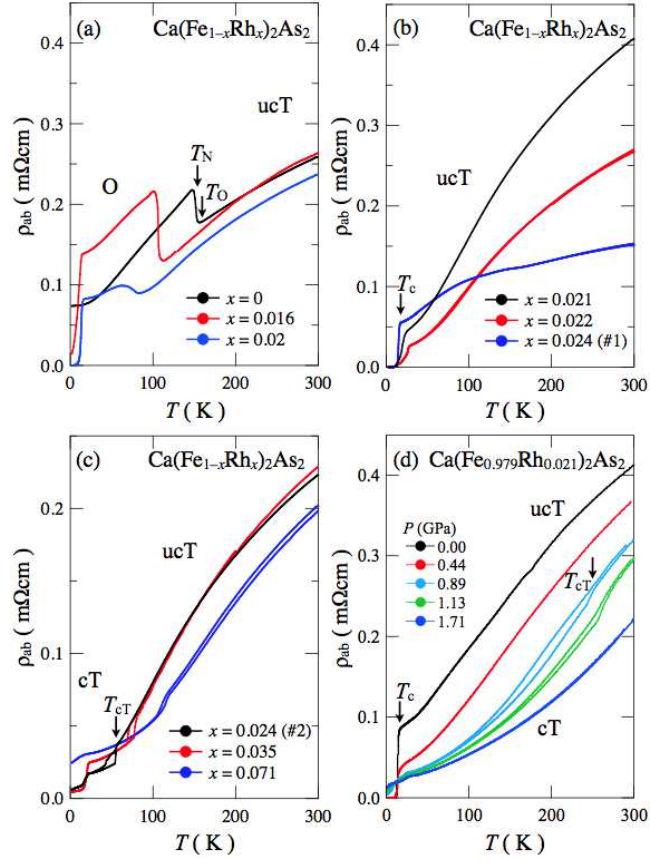}
\caption{(Color online) Temperature dependence of the in-plane resistivity  $\rho_{ab}$ for Ca(Fe$_{1-x}$Rh$_x$)$_2$As$_2$ (a)-(c) with various $x$ under ambient pressure and (d) with $x$ $=$ 0.021 under hydrostatic pressures. 
The $\rho_{ab}$ in (c) and (d) was measured upon heating and cooling. 
The O, ucT, and cT denote the orthorhombic, uncollapsed-tetragonal, and collapsed-tetragonal structures, respectively. 
$T_{\rm O}$, $T_{\rm N}$, $T_{\rm c}$ and $T_{\rm cT}$ are ucT-O, AF, superconducting, and ucT-cT transition temperatures, respectively. 
} 
\end{center}
\end{figure}
\begin{figure}[t]
\begin{center}
\includegraphics[width=8.5cm]{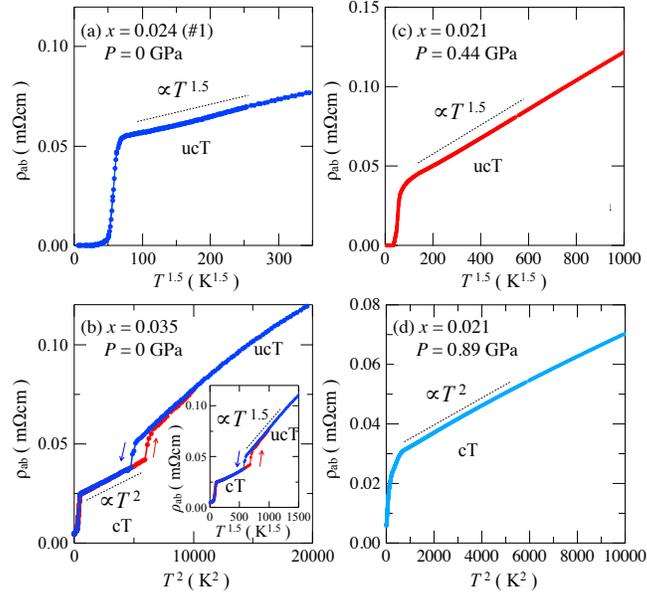}
\caption{(Color online) $\rho_{ab}$ vs $T^n$ plot for Ca(Fe$_{1-x}$Rh$_x$)$_2$As$_2$ with (a) $x$ $=$ 0.024 (\#1) and (b) 0.035 under ambient pressure; and with $x$ $=$ 0.021 under hydrostatic pressures of (c) 0.44 GPa and (d) 0.89 GPa. 
O, ucT, and cT denote the orthorhombic, uncollapsed-tetragonal, and collapsed-tetragonal structures, respectively. 
Broken lines are guides for eyes. 
} 
\end{center}
\end{figure}
\begin{figure}[h]
\begin{center}
\includegraphics[width=8.5cm]{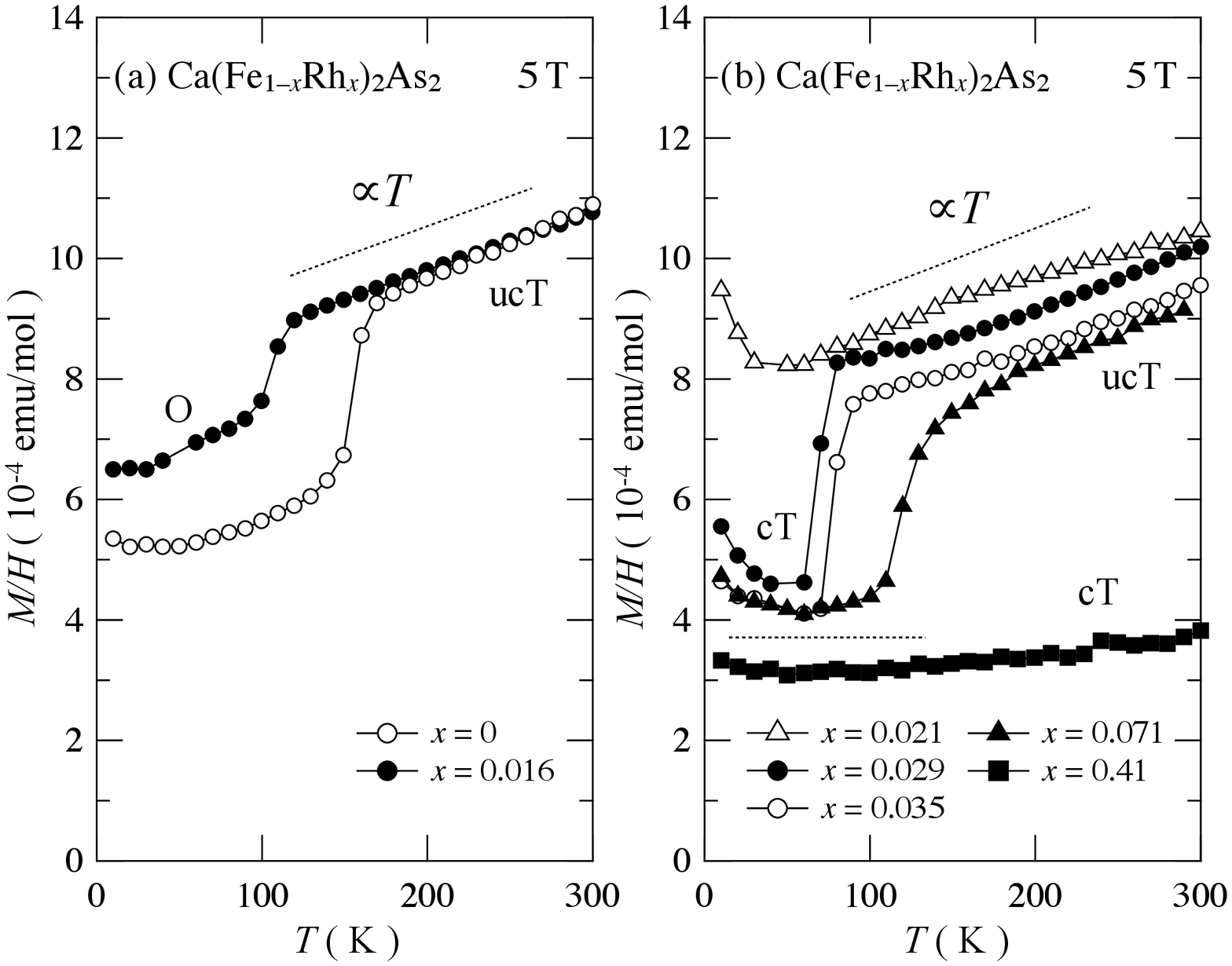}
\caption{Temperature dependence of magnetization $M$ divided by $H$, $M/H$, in a magnetic field of 5 T for Ca(Fe$_{1-x}$Rh$_x$)$_2$As$_2$ with (a) $x$ $=$ 0 and 0.016; and (b) $x$ $=$ 0.021, 0.029, 0.035, 0.071, and 0.41 upon heating. 
O, ucT, and cT denote the orthorhombic, uncollapsed-tetragonal, and collapsed-tetragonal phases, respectively. 
Broken lines are guides for eyes. 
} 
\end{center}
\end{figure}

Figure 1 shows the electronic phase diagram of Ca(Fe$_{1-x}$Rh$_x$)$_2$As$_2$ derived from the results of electrical resistivity, magnetization, and X-ray diffraction. 
The system exhibits a successive phase transition in the $T$ $=$ 0 limit upon Rh doping $x$, and experiences a lattice collapse transition,  analogous to that of CaFe$_2$(As$_{1-x}$P$_x$)$_2$\cite{rf:Kasahara}. 
At 0.00 $\le$ $x$ $\le$ 0.020, the appearance of the antiferromagnetic (AF) phase with orthorhombic (O) structure can be seen in the resistive anomaly at elevated temperatures, as shown in Fig. 2(a). 
We assigned the ucT-O and AF transition temperatures $T_{\rm O}$ and $T_{\rm N}$ as the temperatures below which $\rho_{ab}$ begins to increase and exhibits an inflection point in $d\rho/dT$, respectively, based on the criteria in Refs. \citen{rf:Harnagea} and \citen{rf:Pratt}. 
At $x$ $=$ 0.020, the system exhibits both superconducting and ucT-O transition very likely due to the phase separation at the critical boundary of ucT-O phases in the $T$ $=$ 0 limit. 
At 0.020 $\le$ $x$ $\le$ 0.024, superconducting (SC) phase in the ucT structure was evidenced from the zero resistivity in Fig. 2(b) and the shielding effect shown in the lower inset of Fig. 1. 
The transition temperature $T_{\rm c}$ was determined (from  the onset of the diamagnetism) to be 12, 13, 13, and 14 K for $x$ $=$ 0.020, 0.021, 0.022, and 0.024, respectively. 
The large shielding volume fraction at 2 K ensures the emergence of bulk superconductivity. 
At $x$ $\geq$ 0.024\cite{rf:x=0024}, the lattice collapse transition was observed as a hysteretic drop of $\rho_{ab}$, a characteristic of the ucT-cT transition\cite{rf:Kasahara,rf:Canfield,rf:Yu}, as can be seen in Fig. 2(c). 
The lattice collapse transition was also observed as a discontinuous shrinkage of $c$ parameter upon Rh doping at $x$ $=$ 0.19 at room temperature, as shown in the upper inset of Fig. 1. 
The superconductivity was completely suppressed along with the ucT-cT transition. 
For this reason, the superconductivity emerges in such a narrow range of doping $x$ in Ca(Fe$_{1-x}$Rh$_x$)$_2$As$_2$ as compared to the other iron-based superconductors \cite{rf:Ishida,rf:Paglione,rf:Johnston}. 
For instance, superconductivity emerges in a wide range of $x$, 0.025 $\le$ $x$ $\le$ 0.18, for Ba(Fe$_{1-x}$Co$_x$)$_2$As$_2$, a system in which no ucT-cT transition is exhibited\cite{rf:Chu}. 
In fact, the ucT-cT transition has a large impact on the electronic state. 
The contour plot in Fig. 1 represents the temperature($T$)-and-doping($x$) evolution of the magnetic susceptibility $M/H$ measured in a magnetic field of 5 T\cite{rf:contour}. 
The large drop in $M/H$ at the ucT-cT phase boundary suggests the strong suppression of the electronic density of states (DOS) at the Fermi energy. 
Using the formula of Pauli paramagnetism, we estimate the reduction in the DOS to be approximately $- 4$ to $- 6$ state/eV/Fe from the drop in $M/H$ at the transition. 
The estimate is comparable to $-4.0$ state/eV/Fe predicted from the band calculation by Yildirim {\it et al}\cite{rf:Yildirim}.

A lattice collapse transition was also observed under hydrostatic pressure in Rh-doped Ca(Fe$_{1-x}$Rh$_x$)$_2$As$_2$ with $x$ $=$ 0.021. 
The ucT-cT transition was evidenced from a hysteretic drop of $\rho_{ab}$ at $P$ $\ge$ 0.89 GPa, as shown in Fig. 2(d). 
Bulk superconductivity was observed so long as the system was in the ucT phase; no bulk superconductivity was observed in the cT phase under hydrostatic pressure.

In this way, the lattice collapse transition can be realized using either chemical doping or the application of hydrostatic pressure. 
Superconductivity disappears in the cT phase of CaFe$_2$As$_2$, irrespective of how the cT phase is realized, {\it i.e.,} under the conditions of isovalent (P) doping \cite{rf:Kasahara}, electron (Rh) doping, or hydrostatic pressure, as demonstrated by the present study.

Kasahara {\it et al.} \cite{rf:Kasahara} have reported the switch of non-Fermi-liquid to Fermi-liquid transport at the ucT-cT transition in CaFe$_2$(As$_{1-x}$P$_x$)$_2$. 
We confirmed that the same kinds of switching also occurs in Ca(Fe$_{1-x}$Rh$_x$)$_2$As$_2$. 
As shown in Fig. 3(a), Ca(Fe$_{1-x}$Rh$_x$)$_2$As$_2$ with $x$ $=$ 0.024 exhibits non-Fermi-liquid transport, $\rho_{ab}$ $\propto$ $T^{1.5}$, in the ucT phase. 
In contrast, Ca(Fe$_{1-x}$Rh$_x$)$_2$As$_2$ with $x$ $=$ 0.035 exhibits Fermi-liquid-transport, $\rho_{ab}$ $\propto$ $T^2$, in the cT phase at low temperatures, as shown in Fig. 3(b), although it shows non-Fermi-liquid transport, $\rho_{ab}$ $\propto$ $T^{1.5}$, in the ucT phase at high temperatures, as shown in the inset of Fig. 3(b). 
Moreover, under a hydrostatic pressure $P$ $=$ 0.44 GPa, Ca(Fe$_{1-x}$Rh$_x$)$_2$As$_2$ with $x$ $=$ 0.021 exhibits non-Fermi-liquid transport in the ucT phase, whereas Fermi-liquid transport is restored in the cT phase at $P$ $=$ 0.89 GPa. 
Thus, the change from non-Fermi-liquid to Fermi-liquid transport universally occurs along with the lattice collapse transition in CaFe$_2$As$_2$.

Non-Fermi-liquid behavior, $\rho_{ab}$ $\propto$ $T^{1.5}$ in CaFe$_2$As$_2$-based superconductors as well as $\rho_{ab}$ $\propto$ $T$ in other iron-based superconductors \cite{rf:Liu1,rf:Gooch,rf:Kasahara2}, are thought to be derived from the AF fluctuation near QCPs\cite{rf:Lohneysen}. 
It is most likely that the magnetic fluctuation, related to the Fermi-surface nesting\cite{rf:Mazin,rf:Kuroki,rf:Ding,rf:Liu,rf:Brouet,rf:Terashima}, is responsible for non-Fermi-liquid transport in iron-based superconductors. 
Thus, the recovery of Fermi-liquid transport in the cT phase can be regarded as the disappearance of the magnetic fluctuation, related to the lifting of the Fermi-surface nesting\cite{rf:Kasahara,rf:Tompsett}.

The temperature dependence of magnetic susceptibility $M/H$ provides further evidence for the presence of magnetic fluctuations in the ucT phase and the disappearance of the fluctuations in the cT phase. 
The magnetic susceptibility of Ca(Fe$_{1-x}$Rh$_x$)$_2$As$_2$ in the ucT phase is characterized by $T$-linear behavior, as shown in Fig. 4. 
This $T$-linear susceptibility in the ucT phase changes into (nearly) $T$-independent behavior in the cT phase. 
 $T$-linear susceptibility is widely observed in the normal state of iron-based superconductors \cite{rf:Lumsden}. 
Zhang {\it et al.}\cite{rf:Zhang} theorized that the $T$-linear behavior originates from short-range AF fluctuations. 
The observed temperature dependence of susceptibility indicates the existence and absence of the magnetic fluctuation in the ucT and cT phases, respectively, in accordance with non-Fermi-liquid and Fermi-liquid transport in the ucT and cT phases, respectively.
The transport and magnetic properties suggest that magnetic fluctuation exists in the ucT phase and vanishes in the cT phase. 
Because the superconductivity suddenly disappears in the cT phase, it is plausible that the magnetic fluctuations play a key role in the emergence of superconductivity in iron-based superconductors.

In summary, we observed the lattice collapse transition in Ca(Fe$_{1-x}$Rh$_x$)$_2$As$_2$ at  $x$ $\geq$ 0.024. 
Application of hydrostatic pressure also resulted in a lattice collapse transition at $P$ $\sim$ 0.5 GPa in  Ca(Fe$_{1-x}$Rh$_x$)$_2$As$_2$ with $x$ $=$ 0.021. 
In both cases, the superconductivity of the ucT phase abruptly disappeared in the cT phase, and, simultaneously, the non-Fermi-liquid transport in the ucT phase switched to Fermi-liquid transport in the cT phase. 
Moreover, we showed that the $T$-linear paramagnetic susceptibility in the ucT phase changed into (nearly) $T$-independent behavior in the cT phase. 
All of these observations point toward the disappearance of magnetic fluctuations in the cT phase triggered by the lattice collapse transitions. 
Our result suggests the important role of magnetic fluctuations in the occurrence of superconductivity in CaFe$_2$As$_2$.

We thank M. Takasuga and Y. Nishikubo for their technical assistance. 
Part of this work was performed at the Advanced Science Research Center, Okayama University. 
This work was partially supported by KAKENHI from JSPS and MEXT, Japan.


\begin{thebibliography}{10}
\bibitem{rf:x=0024}
At $x$ $=$ 0.024, we observed different properties depending on batches: sample \#1 remains in the ucT phase at low temperatures as shown in Fig. 1(b), while sample \#2 shows ucT-cT transition as can be seen from Fig. 2(c).
\bibitem{rf:contour}
The contour plot was made from $M/H$-vs-$T$ data for $x$ $=$ 0, 0.016, 0.021, 0.029, 0.035, 0.071, 0.12, 0.19, and 0.41. 
Data at the selected $x$ are shown in Fig. 4 for the sake of clarity.
\end{thebibliography}

\providecommand{\noopsort}[1]{}\providecommand{\singleletter}[1]{#1}%

\end{document}